# INFLUENCE OF RANDOM IRREGULARITIES ON QUASI-THERMAL NOISE SPECTRUM OF PLASMA

**Nikolay A. Zabotin** [†], **Yury V. Chugunov** [‡], **Evgene A. Mareev** [‡], **Andrey G. Bronin** [†]

[†] *Rostov State University, Rostov-on-Don, Russia*
[‡] *Institute of Applied Physics RAS, Nizhni Novgorod, Russia*

## Introduction

In the past three decades the thermal noise spectroscopy was recognized as a fruitful tool of space plasma diagnostics. It is well-known that when a passive electric antenna is immersed into a stable plasma, the thermal motion of the ambient electrons and ions produces fluctuations of the electric potential at the antenna terminals [*Rostoker, 1961; Andronov, 1966; De Passiz, 1969; Fejer and Kan, 1969*]. This quasi-thermal noise may be calculated if the particle velocity distribution function is known [*Rostoker, 1961*]. Since the noise spectrum depends on the main characteristics of plasma, as electron density or bulk temperature, the quasi-thermal noise spectroscopy can be used for diagnostics of plasma parameters. This diagnostic method is most appropriate for in situ space plasma measurements because it does not require of additional apparatus; large scale space plasma conditions allow one to construct antennas whose impedance is very small as compared with input impedance and whose characteristics can be calculated with fair accuracy. Such antennas permit direct observation of the frequency spectrum of thermal noise. Some examples of application of this diagnostic method can be found, for example, in [*Meyer-Vernet, 1979; Couturier et al., 1981; Kellog, 1981; Meyer-Vernet and Pershe, 1989*].

It is also well known that random irregularities of electron density always present in real space plasma. Random irregularities of the Earth's ionosphere are studied intensively and main properties of their spatial spectrum are known [*Fejer and Kelley, 1980; Szuszczewicz, 1986*]. These irregularities considerably affect propagation of radio waves in space plasma changing their phase, amplitude, spatial and angular distribution [*Zabotin, Bronin and Zhbankov, 1998*], as well as group propagation time and pulse duration [*Bronin, Zabotin and Kovalenko, 1999*]. Some information about the irregularity spectra of solar wind is also available [*Rickett, 1973*]. For the purposes of present investigation it is possible to use twin models of shape of spatial spectrum for both ionospheric and solar wind irregularities. Though that does not relate to parameters of the spectra.

Irregularities substantially change properties of the medium with relation to electromagnetic radiation and they may also influence quasi-thermal noise spectrum detected by antenna. What is the possible physical mechanism of this influence? It is known, that fluctuations in plasma are closely connected with plasma dissipative properties. From the viewpoint of statistical mechanics, random irregularities in plasma may be understood as non-thermal large scale (in comparison with characteristic time and scale of particle motion) fluctuations. Such fluctuations may considerably change collision term in kinetic equation [*Klimontovich, 1982*] and, consequently, velocity distribution function. Since the quasi-thermal noise spectrum is determined by velocity distribution function, the change in distribution function will lead to change in noise spectrum. From the viewpoint of electrodynamics, random irregularities change mean dielectric properties of the medium (see, for example, [*Hagfors, 1984*]). It means that in the media with random irregularities the roots of dispersion equation are shifted with relation to undisturbed values and even new roots may appear. The imaginary part of these roots may also be considerably changed due to additional non-collisional attenuation caused by scattering of waves in random medium.

Development of strict theory of influence of random irregularities on quasi-thermal noise spectrum is a very complicated and difficult task, involving various fields of electrodynamics and statistical physics. In the present paper we will assume that noise spectrum in the random medium may be calculated using the tensor of effective dielectric permittivity. This tensor is determined as the dielectric permittivity tensor of some imaginary "effective" regular medium in which the field of point source is the same as the mean field in the corresponding random medium [*Ryzhov, Tamoikin and Tatarskii, 1965, Ryzhov 1967, Ryzhov 1968*]. Since effective dielectric permittivity tensor determines the mean field of the source in random media it also determines the impedance of antenna, which in its turn, determines the noise spectrum. It has been shown also that correlation function of electric field fluctuations in random medium may be expressed through the effective dielectric constant tensor [*Ryzhov, 1968*]. (It should be noted, however, that this result is based on averaging of Kallen-Welton formula and is valid only for states near the thermal equilibrium.) Spectrum of electrostatic noise in its turn is completely determined by correlation function of longitudinal electric field [*Meyer-Vernet and Pershe, 1989*]. The method based on effective dielectric constant tensor was used to study thermal noise spectrum in dielectrics [*Ryzhov, 1970*]. We apply this approach to the irregular space plasma.

Calculation of the effective dielectric permittivity tensor of plasma is also rather complex problem involving summation of infinite series of perturbation theory. Some approximation is necessary to get the tensor components in closed form. In this paper we follow Born approximation.

Thus, in the present paper we intend to estimate possible changes in the thermal noise frequency spectrum in plasma with random irregularities rather then to develop complete and strict theory of this effect. In Section 1 of the paper we briefly discuss calculation of the noise spectrum and impedance of antenna in Maxwellian plasma (we assume that presence of irregularities does not change velocity distribution function radically and it may be approximately described by Maxwell distribution). In Section 2 the method of calculation of effective dielectric permittivity tensor is stated. Section 3 discusses specific for given problem difficulties in numerical calculations. Results of numerical calculations in application to ionospheric plasma and plasma of solar wind are collected in Section 4.

## 1. Quasi-thermal noise spectrum and antenna impedance

Usually the noise voltage spectral density measured at the electric antenna terminals, which is immersed in a plasma may be expressed through the spatial Fourier transform of the current distribution in the antenna and autocorrelation function of electrostatic field fluctuations in the antenna frame [*Rostoker, 1961*]. The shot noise, drift of the plasma across the antenna as well as some other phenomena also contribute into noise spectra. We do not take these phenomena into account now. If the plasma is in the thermal equilibrium at temperature $T$, what will be assumed in further consideration, the voltage spectral density may be expressed through antenna impedance by the formula [*Meyer-Vernet and Pershe, 1989*]:

$$V^2 = 4k_B T \, \text{Re}(Z), \qquad (1)$$

where $k_B$ is Boltzmann constant, $Z$ is antenna impedance and Re denotes the real part.

Calculation of antenna impedance in general case is rather complex problem. One must take into account various phenomena such as the disturbance of trajectories of particles, collection of electrons and ions, photoemission and so on. The common approximation is to take into account only electron plasma oscillations. In this approximation one has

$$\text{Re}(Z) = \frac{1}{(2\pi)^3 \varepsilon_0 \omega} \int d^3k \, \frac{\text{Im}\,\varepsilon_\|(\omega, \vec{k})}{\left|\varepsilon_\|(\omega, \vec{k})\right|^2} \cdot \frac{\left|\vec{k} \cdot \vec{J}(\vec{k})\right|^2}{k^2}, \qquad (2)$$

where $\varepsilon_\|(\omega, \vec{k})$ is the plasma longitudinal permittivity, $\vec{J}(\vec{k})$ is the spatial Fourier transform of the antenna current distribution. For Maxwellian collisionless plasma longitudinal dielectric permittivity is given by the well-known expression [*Akhiezer, 1972*]:

$$\varepsilon_\|(\omega, \vec{k}) = 1 + \frac{k_D^2}{k^2}\left[1 - \Phi(z) + i\pi^{1/2} z \exp(-z^2)\right], \qquad (3)$$

where $\Phi(z) = 2z\exp(-z^2)\int_0^z \exp(t^2)dt$, $z = \omega k_D/(\sqrt{2}\omega_p k)$, $k_D = 1/r_D$, $r_D$ is Debye's length, $\omega_p$ is plasma frequency.

To calculate impedance one needs to choose the current distribution $\vec{J}(\vec{k})$. The frequently used model of current distribution corresponds to wire dipole antenna [*Kuhl 1966; Kuhl 1967; Kellog 1981*]. The geometry of wire dipole antenna is shown in Fig.1. The antenna consists of two cylinders, each of length $L$ and radius $a << L$. This model may be a good approximation for geometry of real antenna used for space missions. The antenna parameters used in our calculations are in close correspondence with real parameters of spacecraft antennae used in investigations of ionosphere or solar wind (Wind and Kosmos-1809 missions).

For wire dipole antenna one can get [*Meyer-Vernet and Pershe, 1989; Kuhl, 1966, Kellog, 1981*]:

$$\mathrm{Re}(Z) = \frac{4}{\pi^2 \varepsilon_0 \omega} \int_0^\infty dk \frac{\mathrm{Im}\,\varepsilon_\|(\omega,k)}{|\varepsilon_\|(\omega,k)|^2} F(k), \quad (3)$$

where

$$F(k) = [J_0(ka)]^2 \frac{\left\{Si(kL) - Si(2kL) - \dfrac{2\sin^4(kl)}{kL}\right\}}{kL},$$

$J_0(x)$ is Bessel function of the first kind and $Si(x)$ is integral sinus function (see, for example, [*Abramowitz and Stegun, 1964*]). To calculate quasi-thermal electromagnetic noise spectrum in the plasma with random irregularities one should substitute the longitudinal dielectric permittivity for regular plasma $\varepsilon_\|(\omega,\vec{k})$ by effective longitudinal dielectric permittivity $\varepsilon_\|^{eff}(\omega,\vec{k})$ [*Ryzhov, 1968*].

## 2. Calculation of effective dielectric permittivity tensor

General scheme of calculation of effective dielectric permittivity tensor starts from the Dyson equation for the mean field in random medium (see [*Rytov, Kravtsov and Tatarskii, 1978*]). It may be easily found that effective dielectric permittivity tensor is proportional to the Fourier transform of the mass operator [*Ryzhov, 1967*]. Taking into account only initial term in the mass operator series expansion (what corresponds to Born approximation for the mean field) one should obtain [*Ryzhov, 1967*]

$$\varepsilon_{ij}^{eff}(\omega,\vec{k}) = \varepsilon_{ij}^{0}(\omega,\vec{k}) - k_0^2 B_{imlj}(\omega,\vec{k}) \int G_{ml}(\omega,\vec{p}) \Phi_N(\vec{k}-\vec{p}) d^3p \,, (4)$$

where $B_{imlj} = (\varepsilon_{im}^0 - \delta_{im})(\varepsilon_{lj}^0 - \delta_{lj})$, $\varepsilon_{ij}^0(\omega,\vec{k})$ is dielectric permittivity tensor of regular plasma, $G_{ml}(\omega,\vec{p})$ is the Fourier transform of Green tensor, $k_0 = \omega/c$, $c$ is velocity of light in vacuum, $\Phi_N(\vec{k})$ is spatial spectrum of irregularities.

In isotropic plasma one has

$$\varepsilon_{ij}^0(\omega,\vec{k}) = \left(\delta_{ij} - \frac{k_i k_j}{k^2}\right)\varepsilon_\perp(\omega,\vec{k}) + \frac{k_i k_j}{k^2}\varepsilon_\parallel(\omega,\vec{k}); (5)$$

$$G_{lm}(\omega,\vec{k}) = \left(\delta_{ij} - \frac{k_i k_j}{k^2}\right)\frac{1}{k_0^2 \varepsilon_\perp(\omega,\vec{k}) - k^2} + \frac{k_i k_j}{k^2}\frac{1}{k_0^2 \varepsilon_\parallel(\omega,\vec{k})}, (6)$$

where longitudinal permittivity $\varepsilon_\parallel(\omega,\vec{k})$ is defined by formula (3) and transversal permittivity is determined by the expression [*Akhiezer et al, 1972*]:

$$\varepsilon_\perp(\omega,\vec{k}) = 1 - \frac{\omega^2}{\omega_p^2}\left(\Phi(z) - i\pi^{1/2} z \exp(-z^2)\right).$$

At this point of our consideration we should choose a concrete type of the irregularity spatial spectrum $\Phi_N(\vec{k})$. Ionospheric irregularities are described by complex spectrum, which properties are different for different intervals of wavenumbers of irregularities, but for the certain interval of wavenumbers the spectrum may be described by power law:

$$\Phi_N(\vec{k}) \propto \left(1 + l_x^2 k_x^2 + l_y^2 k_y^2 + l_z^2 k_z^2\right)^{-\mu/2}.$$

Generally the spectrum is anisotropic: irregularities may be strongly stretched along the lines of force of geomagnetic field due to difference in the diffusion coefficients for longitudinal and transversal directions. However, for simplicity of calculations, we will use the model of isotropic spectrum. This situation is less characteristic of topside ionospheric plasma, but may be considered normal for solar wind plasma.

In isotropic medium with isotropic irregularities tensor of effective dielectric permittivity has the same structure as tensor in regular media (5) (in other words effective medium is also isotropic):

$$\varepsilon_{ij}^{eff}(\omega,\vec{k}) = \left(\delta_{ij} - \frac{k_i k_j}{k^2}\right)\varepsilon_\perp^{eff}(\omega,\vec{k}) + \frac{k_i k_j}{k^2}\varepsilon_\parallel^{eff}(\omega,\vec{k}), (7)$$

where

$$\varepsilon_{\parallel}^{eff}(\omega,\vec{k}) = \varepsilon_{\parallel}(\omega,\vec{k}) - [\varepsilon_{\parallel}(\omega,\vec{k})-1]^2 \left\{ \int \frac{(\vec{k}\cdot\vec{p})}{k^2 p^2} \frac{\Phi_N(\vec{k}-\vec{p})}{\varepsilon_{\parallel}(\omega,\vec{k})} d^3 p + \right.$$

$$\left. k_0^2 \int \left[1 - \frac{(\vec{k}\cdot\vec{p})}{k^2 p^2}\right] \frac{\Phi_N(\vec{k}-\vec{p})}{k_0^2 \varepsilon_{\perp}(\omega,\vec{k}) - p^2} d^3 p \right\}; \quad (8)$$

$$\varepsilon_{\perp}^{eff}(\omega,\vec{k}) = \varepsilon_{\perp}(\omega,\vec{k}) - \frac{1}{2}[\varepsilon_{\perp}(\omega,\vec{k})-1]^2 \left\{ \int \left[1 - \frac{(\vec{k}\cdot\vec{p})}{k^2 p^2}\right] \frac{\Phi_N(\vec{k}-\vec{p})}{\varepsilon_{\parallel}(\omega,\vec{k})} d^3 p + \right.$$

$$\left. k_0^2 \int \left[1 + \frac{(\vec{k}\cdot\vec{p})}{k^2 p^2}\right] \frac{\Phi_N(\vec{k}-\vec{p})}{k_0^2 \varepsilon_{\perp}(\omega,\vec{k}) - p^2} d^3 p \right\}. \quad (9)$$

For specific spectrum index $\mu = 4$ (quite typical value both for ionospheric and for solar wind plasma) the spectrum can be written as follows:

$$\Phi_N(\vec{k}) = \frac{\delta_R^2 l^3}{2\pi^2 (1-\exp(-R/l))} (1+l^2 k^2)^{-2}, \quad (10)$$

where $l = L_m / 2\pi$, $L_m$ is the outer scale of spectrum, $\delta_R^2 = D(R)$, $D(R)$ is structure function of irregularities at scale length $R$. Substituting (10) into (8) one obtains for longitudinal effective dielectric permittivity:

$$\varepsilon_{\parallel}^{eff}(\omega,\vec{k}) = \varepsilon_{\parallel}(\omega,\vec{k}) - \frac{\delta_R^2 l^3}{\pi(1-\exp(-R/l))} [\varepsilon_{\parallel}(\omega,\vec{k})-1]^2 \times$$

$$\left\{ \int_0^{\infty} \frac{p^2 dp}{\varepsilon_{\parallel}(\omega,\vec{p})} \int_{-1}^{1} \frac{t^2 dt}{(1+l^2 k^2 + l^2 p^2 - 2l^2 kpt)^2} + \right. \quad (11)$$

$$\left. k_0^2 \int_0^{\infty} \frac{p^2 dp}{k_0^2 \varepsilon_{\perp}(\omega,\vec{p}) - p^2} \int_{-1}^{1} \frac{(1-t^2) dt}{(1+l^2 k^2 + l^2 p^2 - 2l^2 kpt)^2} \right\}.$$

Internal integration in (11) can be done in analytical form:

$$\int_{-1}^{1} \frac{t^2 dt}{(1+l^2 k^2 + l^2 p^2 - 2l^2 kpt)^2} = \frac{1}{2l^2 kp} \left\{ \frac{\xi^2}{1-\xi^2} - \frac{\ln\left|\frac{1+\xi}{1-\xi}\right| - 2\xi}{\xi} \right\};$$

$$\int_{-1}^{1} \frac{(1-t^2) dt}{(1+l^2 k^2 + l^2 p^2 - 2l^2 kpt)^2} = \frac{1}{2l^2 kp} \frac{\ln\left|\frac{1+\xi}{1-\xi}\right| - 2\xi}{\xi}, \quad (12)$$

$$\xi = \frac{2l^2 pk}{1+l^2 k^2 + l^2 p^2}.$$

These functions turn into zero when $\xi$ is zero (what corresponds to limiting cases $p \to 0, p \to \infty, k \to 0$ or $k \to \infty$) and provide proper convergence of integrals in (11) at both limits of integration over $p$.

Though one could use result (11) directly, without substitution of analytic expressions for the inner integral, generally such approach may be not successful. The reason is that integrands in (11) have several peculiarities, which will be discussed below, as well as significant dependence of integrand on such parameters as Debye's radius, outer scale of the irregularity spectrum, dimensions of antenna etc. The latter difficulty, however, may be considerably reduced by introduction of dimensionless variables.

## 3. Details of numerical calculations

Considerable difficulties under numerical calculation of quasi-thermal noise spectrum and of effective dielectric permittivity tensor using expressions (3) and (11) correspondingly are caused by the "bad" behavior of integrand at some special points. In both cases integrand contains expressions of one of the following kinds:

$$\frac{\operatorname{Im} \varepsilon_{\|}(\omega, \vec{k})}{\left(\operatorname{Re} \varepsilon_{\|}(\omega, \vec{k})\right)^2 + \left(\operatorname{Im} \varepsilon_{\|}(\omega, \vec{k})\right)^2} ; (12)$$

$$\frac{\operatorname{Re} \varepsilon_{\|}(\omega, \vec{k})}{\left(\operatorname{Re} \varepsilon_{\|}(\omega, \vec{k})\right)^2 + \left(\operatorname{Im} \varepsilon_{\|}(\omega, \vec{k})\right)^2} ; (13)$$

$$\frac{\operatorname{Im} \varepsilon_{\|}^{eff}(\omega, \vec{k})}{\left(\operatorname{Re} \varepsilon_{\|}^{eff}(\omega, \vec{k})\right)^2 + \left(\operatorname{Im} \varepsilon_{\|}^{eff}(\omega, \vec{k})\right)^2} . (14)$$

As an example, in Fig. 2 the dependence of $\operatorname{Re} \varepsilon_{\|}$ (solid lines) and $\operatorname{Im} \varepsilon_{\|}$ (dashed lines) on $kr_D$ for four values of dimensionless frequency $\omega/\omega_p$ (=1.01; 1.05; 1.1; 1.2) has been plotted. Functions (12) and (14) have a peak of height $\sim 1/\operatorname{Im} \varepsilon_{\|}$ and width $\sim \operatorname{Im} \varepsilon_{\|}$ (or $\sim 1/\operatorname{Im} \varepsilon_{\|}^{eff}$ and $\sim \operatorname{Im} \varepsilon_{\|}^{eff}$) when $\operatorname{Re} \varepsilon_{\|} = 0$ or $\operatorname{Re} \varepsilon_{\|}^{eff} = 0$. For $\omega \sim \omega_p$ the values of $\operatorname{Im} \varepsilon_{\|}$ are very small when $\operatorname{Re} \varepsilon_{\|} = 0$ and the peak is very sharp (see Figs. 4 and 5). Function (13) has two peaks of opposite signs at the left and right of the point where $\operatorname{Re} \varepsilon_{\|} = 0$, and these peaks are also very sharp at the frequencies close to plasma frequency (see Fig. 3). A usual way to calculate integrals with such

quasi-singular integrand is to split the integration interval at the point where the singularity happens. In opposite case the result would depend upon the position of the peak with relation to the grid of abscissas of integration rule which would be different for different frequencies. In other words, in this case the calculated noise spectra would contain random error component due to accumulation of inaccuracy under numerical integration. However the splitting of integration interval is not sufficient for successful integration. Additional difficulties are caused by the fact that the roots of equation $\operatorname{Re}\varepsilon_\| = 0$ may be found only approximately ($\operatorname{Re}\varepsilon_\| \sim 10^{-15}$ in the root), while the magnitude of $\operatorname{Im}\varepsilon_\|$ in the root may be several orders of magnitude smaller. To avoid this interference one should take into account the fact that in real plasma some small collisional attenuation is always present. The value of collision frequency may be chosen so that inequality $\operatorname{Re}\varepsilon_\| \ll \operatorname{Im}\varepsilon_\|$ always holds in the small vicinity of the root of equation $\operatorname{Re}\varepsilon_\| = 0$.

As it has been noticed above it is useful to introduce dimensionless variables to reduce the dependence of numerical calculations on certain values of plasma parameters. It is natural to determine dimensionless frequency $\tilde{\omega}$ and wavenumber $\tilde{k}$ as

$$k = k_0 \tilde{k}, \qquad \omega = \omega_p \tilde{\omega}. \quad (13)$$

In these variables longitudinal dielectric permittivity, for example, may be written as:

$$\varepsilon_\|(\tilde{\omega}, \tilde{k}) = 1 + \frac{1}{(k_0 r_D)^2 \tilde{k}^2}\left[1 - \Phi(z) + i\sqrt{\pi} z \exp(-z^2)\right], (14)$$

where $z = \dfrac{1}{\sqrt{2}} \dfrac{\tilde{\omega}}{k_0 r_D \tilde{k}}$. In (14) plasma properties are taken into account only through the dimensionless constant $k_0 r_D$.

Using (12) – (14), expression (11) for effective dielectric constant tensor can be transformed to the following form

$$\varepsilon_\|^{eff}(\tilde{\omega}, \tilde{k}) = \varepsilon_\|^0(\tilde{\omega}, \tilde{k}) - \left[\varepsilon_\|^0(\tilde{\omega}, \tilde{k}) - 1\right]^2 \frac{2}{\pi} \frac{\delta_R^2}{1-\exp(-R/l)} (k_0 l)^3 \cdot$$

$$\left\{\int_0^\infty d\tilde{p} \frac{\tilde{p}^2}{\left[1+(k_0 l)^2(\tilde{p}-\tilde{k})^2\right]\left[1+(k_0 l)^2(\tilde{p}+\tilde{k})^2\right]} \left(\varepsilon_\|^0(\tilde{\omega}, \tilde{p})\right)^{-1} + \frac{1}{4(k_0 l)^6 \tilde{k}^3}\int_0^\infty d\tilde{p}\frac{\left[1+(k_0 l)^2(\tilde{k}^2+\tilde{p}^2)\right]}{\tilde{p}} \cdot \right.$$

$$\left. \left[\frac{2(k_0 l)^2 \tilde{k}\tilde{p}}{1+(k_0 l)^2(\tilde{k}^2+\tilde{p}^2)} - \frac{1}{2}\ln\left|\frac{1+(k_0 l)^2(\tilde{k}+\tilde{p})^2}{1+(k_0 l)^2(\tilde{k}-\tilde{p})^2}\right|\right]\left(\varepsilon_\|^0(\tilde{\omega}, \tilde{p})\right)^{-1}\right\}. (15)$$

It is easy to show, that the second integral in (15) is at least $(k_0 l)^2$ times smaller than first and since may be neglected.

Integrand of first integral in (15) contains factor $\tilde{p}^2 / \left[1 + (k_0 l)^2 (\tilde{p} - \tilde{k})^2\right]$ which also has singular-like behavior when $k_0 l \gg 1$. This problem may be solved by appropriate substitution of independent variable. For example, the substitution $\tilde{p} = \tilde{k} + \tan\vartheta / k_0 l$ transform integrand into smooth function without singularities at $\tilde{p} = \tilde{k}$.

We used Brent algorithm for searching the roots of equation $\text{Re}\,\varepsilon_\parallel(\tilde{\omega}, \tilde{k}) = 0$; the adaptive integration method based on Gauss-Kronrod rule was used to perform numeric integration in expression (15).

## 4. Results of numerical calculations

For calculations of quasi-thermal noise spectrum we first chose the regular parameters of plasma and spectrum of irregularities corresponding to the ionosphere F-region: plasma frequency $\omega_p / 2\pi$ was equal to 3 MHz, Debye's length – 5 cm; $R = 1$ km, $l = 10/2\pi$ km. Calculations were carried out for four different values of $\delta_R$: $5 \cdot 10^{-3}$, $10^{-2}$, $2 \cdot 10^{-2}$ and $3 \cdot 10^{-2}$. The first value is typical for undisturbed mid-latitude ionosphere, the last is observed in polar and equatorial ionosphere and in the experiments of plasma heating by the powerful HF radiowave.

In regular plasma the spectrum has a peak just above the plasma frequency; the shift from the plasma frequency is of order of $\delta\omega/\omega \sim (r_D/L)^2$ [*Meyer-Vernet and Pershe, 1989*]. For the chosen parameters of antenna $r_D/L \ll 1$ and $r_D/a \ll 1$. It means that noise spectrum does not depend upon the radius of antenna wire and have a sharp peak near the plasma frequency. The dimensionless spectra $\dfrac{\pi^2 \varepsilon_0 \omega}{16 k_B T} V^2$ for regular plasma and for irregular plasma with given $\delta_R$ are plotted in Fig. 6.

In presence of irregularities we observe noticeable change in the spectrum of quasi-thermal noise. The peak of the spectrum in this case is split into two peaks. The splitting of the peak is caused by the complication of dielectric properties of plasma in the presence of random irregularities. At the relatively high magnitudes of $\delta_R$ real part of effective longitudinal dielectric constant has additional roots, while the contribution of the roots existing in regular plasma is dumped due to increase of the imaginary part caused by scattering. The distance between peaks is greater for greater irregularity level. Its magnitude is small, approximately 10 – 15 KHz for $\delta_R = 5 \cdot 10^{-3}$ and ~ 100 KHz for $\delta_R = 3 \cdot 10^{-2}$. Such details of the spectrum, however, can be detected in experiment.

For the solar wind plasma we chose the following parameters: plasma frequency $\omega_p / 2\pi = 20$ KHz, $r_D = 10$ km, $l = 10^6 / 2\pi$ km. Noticeable influence of random irregularities on the electromagnetic noise spectrum takes place at relatively high level of irregularities. In Fig. 7 the noise spectrum is plotted for $\delta_R = 5 \cdot 10^{-2}$, $15 \cdot 10^{-2}$ and $30 \cdot 10^{-2}$ for normalization scale length $R = 10^5$ m. The dimensionless noise spectrum for $\delta_R = 15 \cdot 10^{-2}$ $30 \cdot 10^{-2}$ and $50 \cdot 10^{-2}$ for normalization scale $R = 10^8$ m is plotted in Fig. 8. In the case of solar wind plasma we observe the formation of plateau in the spectrum in the vicinity of plasma frequency instead of the splitting of the peak. This difference is explained by the difference in ratio of Debye's length to wire length for ionospheric plasma and for solar wind plasma.

## Conclusions

In this paper we have considered the influence of random irregularities of electron density in isotropic plasma on the quasi-thermal noise spectrum using fairly simple model of irregularities. We have found that for the small values of irregularity level modification of noise spectrum is negligibly small. However, for larger values of $\Delta N / N$, also quite possible in natural conditions, irregularities cause some noticeable effects. In the ionospheric plasma it is the splitting of the peak in the frequency noise spectrum located just above the plasma frequency, into two peaks. Though the gap between those peaks is small it still may be detected in experiment. The magnitude of the gap depends upon the value of $\Delta N / N$, what makes possible using measurements of noise spectra for the purpose of the irregularity diagnostics. In the solar wind plasma irregularities cause changes of the shape of the main spectrum maximum near the plasma frequency resulting in appearance of the plateau under higher irregularity level. This effect also can provide essential information about the solar wind irregularities.

In this paper we used simplified model of plasma which may be significantly improved. The major possible improvement concerns the spectrum of irregularities. For example, in real topside ionosphere irregularities are stretched along the lines of force of geomagnetic field and their spectrum is anisotropic. One can also take into account drift of the plasma across the antenna, because noise spectra measurement are done on the board of satellite moving through the plasma. Though the account for magnetic field is considered to be unimportant for calculation of noise spectrum in regular plasma, in the presence of irregularities its influence on wave propagation may be important.

For other kinds of space plasma, like the plasma of solar wind, the account of all these factors may be essential, first of all because such plasma may be anisotropic even in the absence of external magnetic field. Besides generally it cannot be considered as being in the thermal equilibrium, so more general approach to deriving expressions for noise spectra may be required.

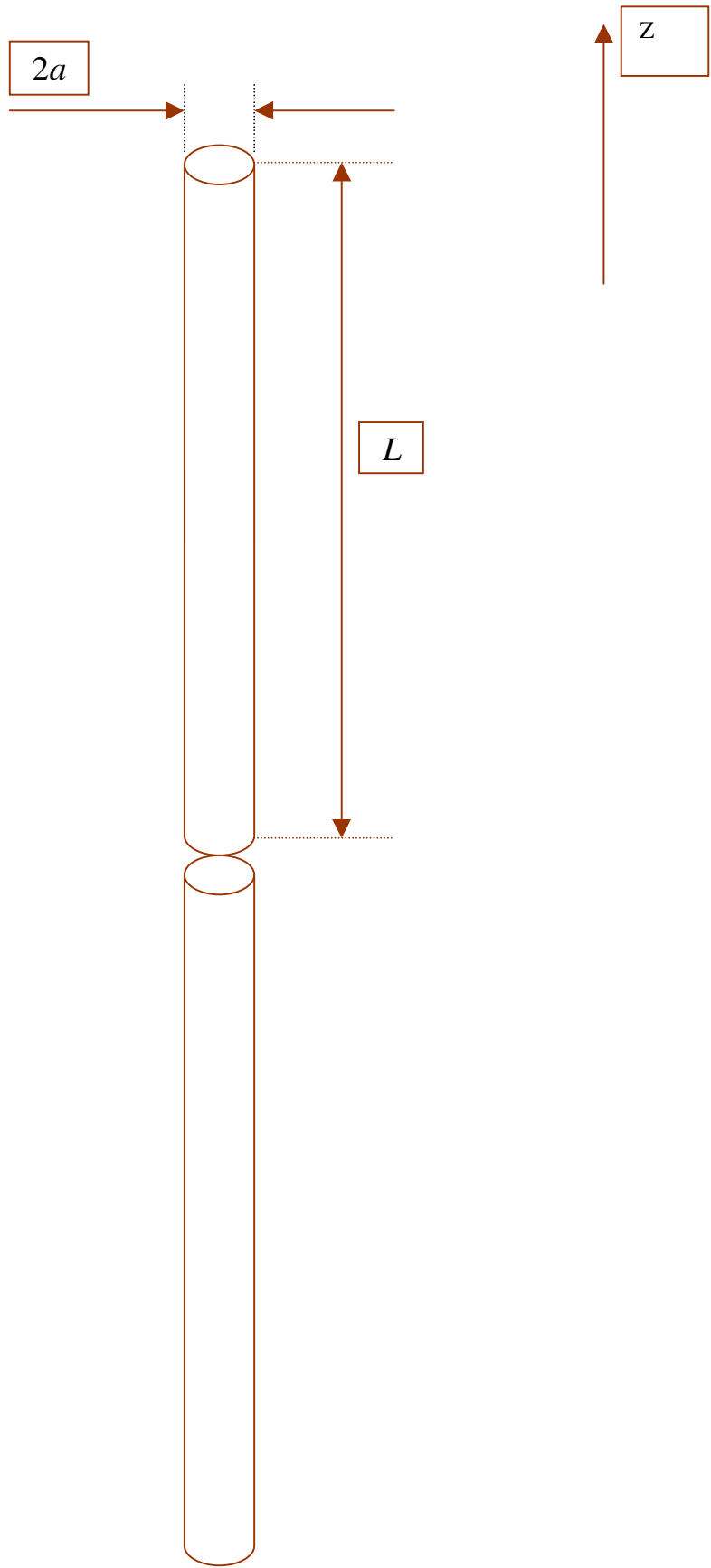

Fig.1.Wiredipoleantennageometry

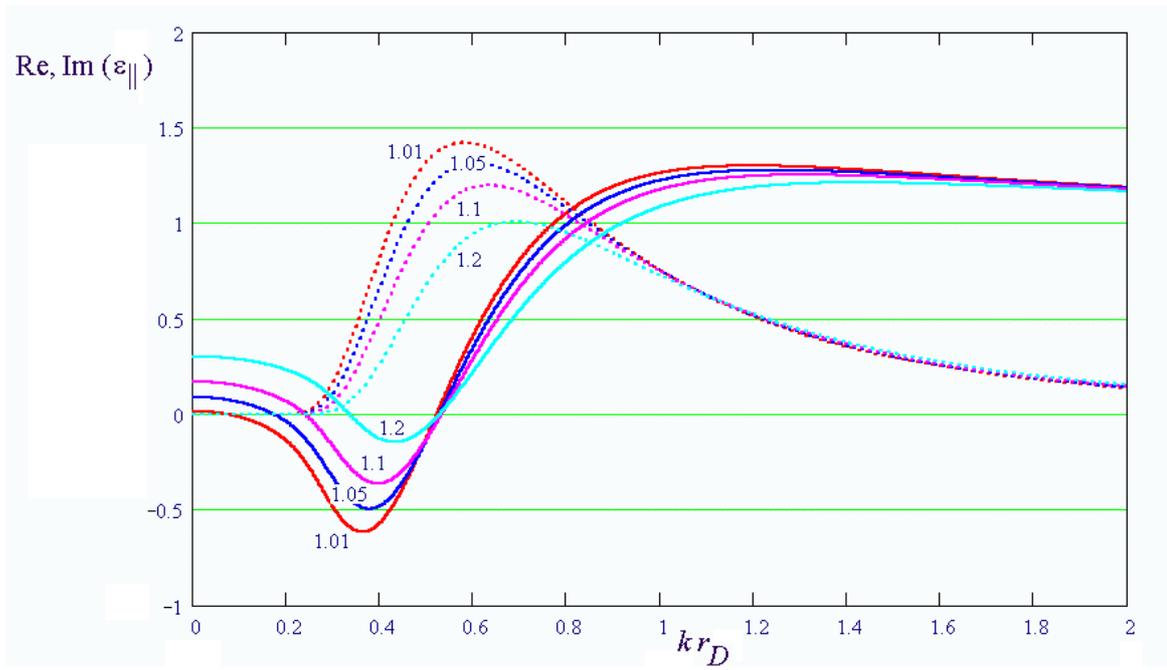

Fig.2. Real and imaginary components of longitudinal dielectric permittivity of a regular plasma at the frequency near plasma frequency. Figures at the curves indicates the ratio of wave frequency to plasma frequency, corresponding to the curve.

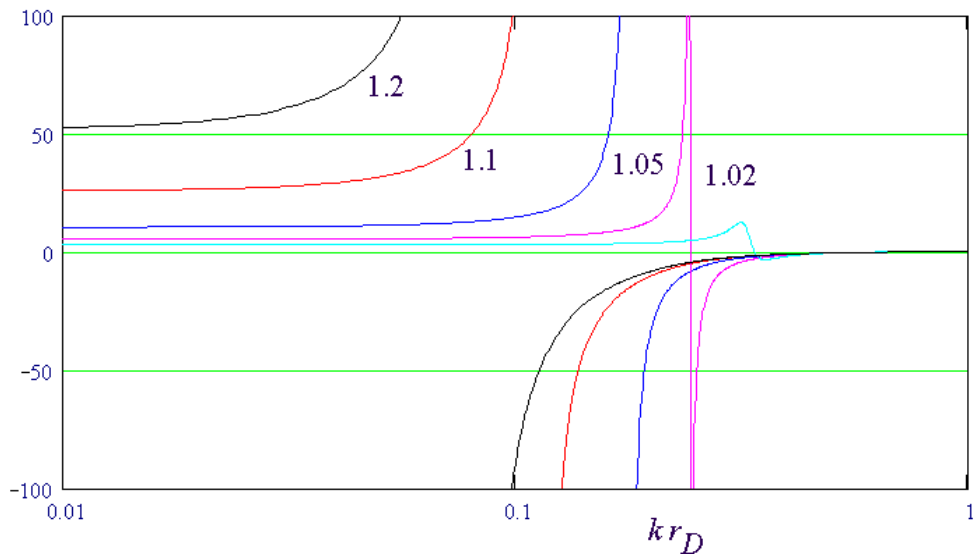

Fig.3. $\dfrac{\operatorname{Re}\varepsilon_{\|}(\omega,\vec{k})}{\left(\operatorname{Re}\varepsilon_{\|}(\omega,\vec{k})\right)^2 + \left(\operatorname{Im}\varepsilon_{\|}(\omega,\vec{k})\right)^2}$ as a function of $k$ in the vicinity of the root of equation $\operatorname{Re}\varepsilon_{\|}=0$.

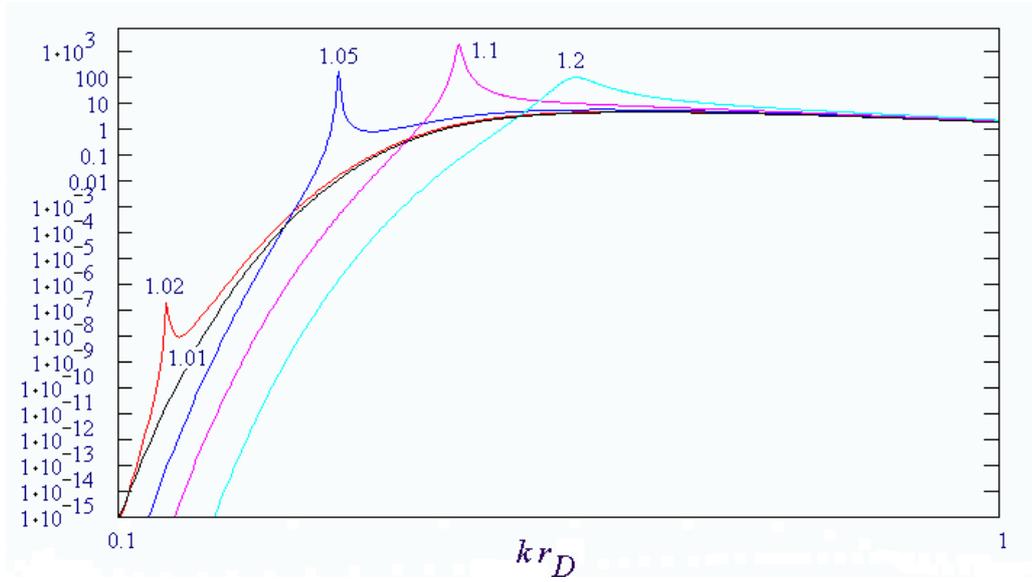

Fig.4. $\dfrac{\operatorname{Im}\varepsilon_{\|}(\omega,\vec{k})}{\left(\operatorname{Re}\varepsilon_{\|}(\omega,\vec{k})\right)^2+\left(\operatorname{Im}\varepsilon_{\|}(\omega,\vec{k})\right)^2}$ as a function of $k$ in the vicinity of the root of equation $\operatorname{Re}\varepsilon_{\|}=0$

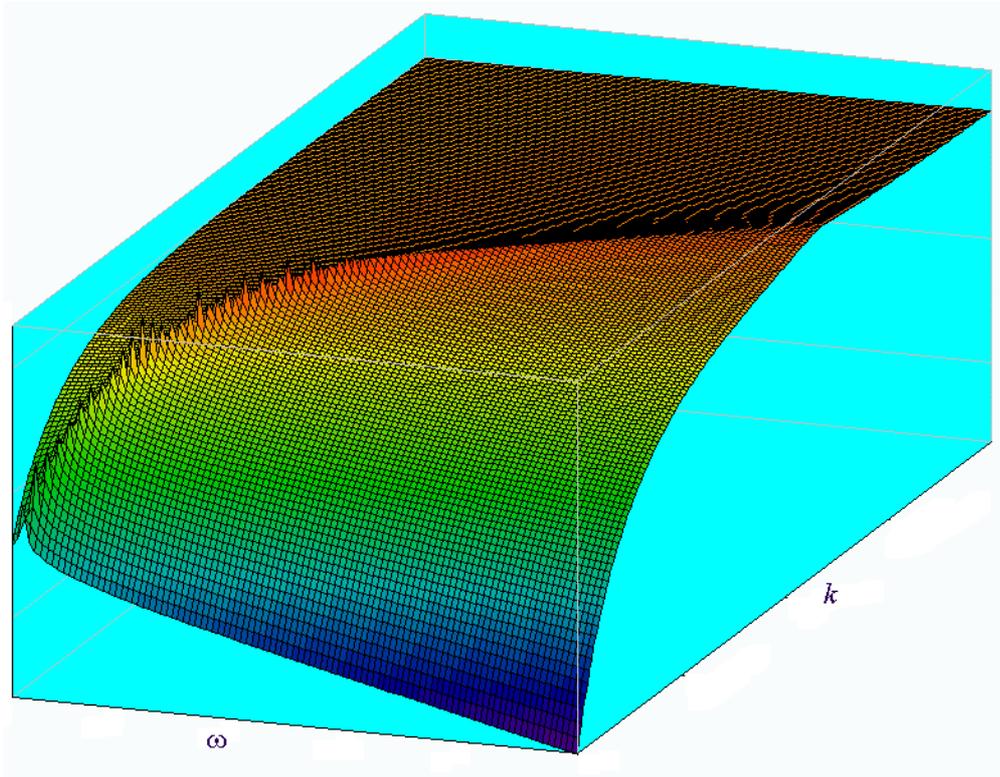

Fig.5. Same as 4, but as function of $k$ and $\omega$.

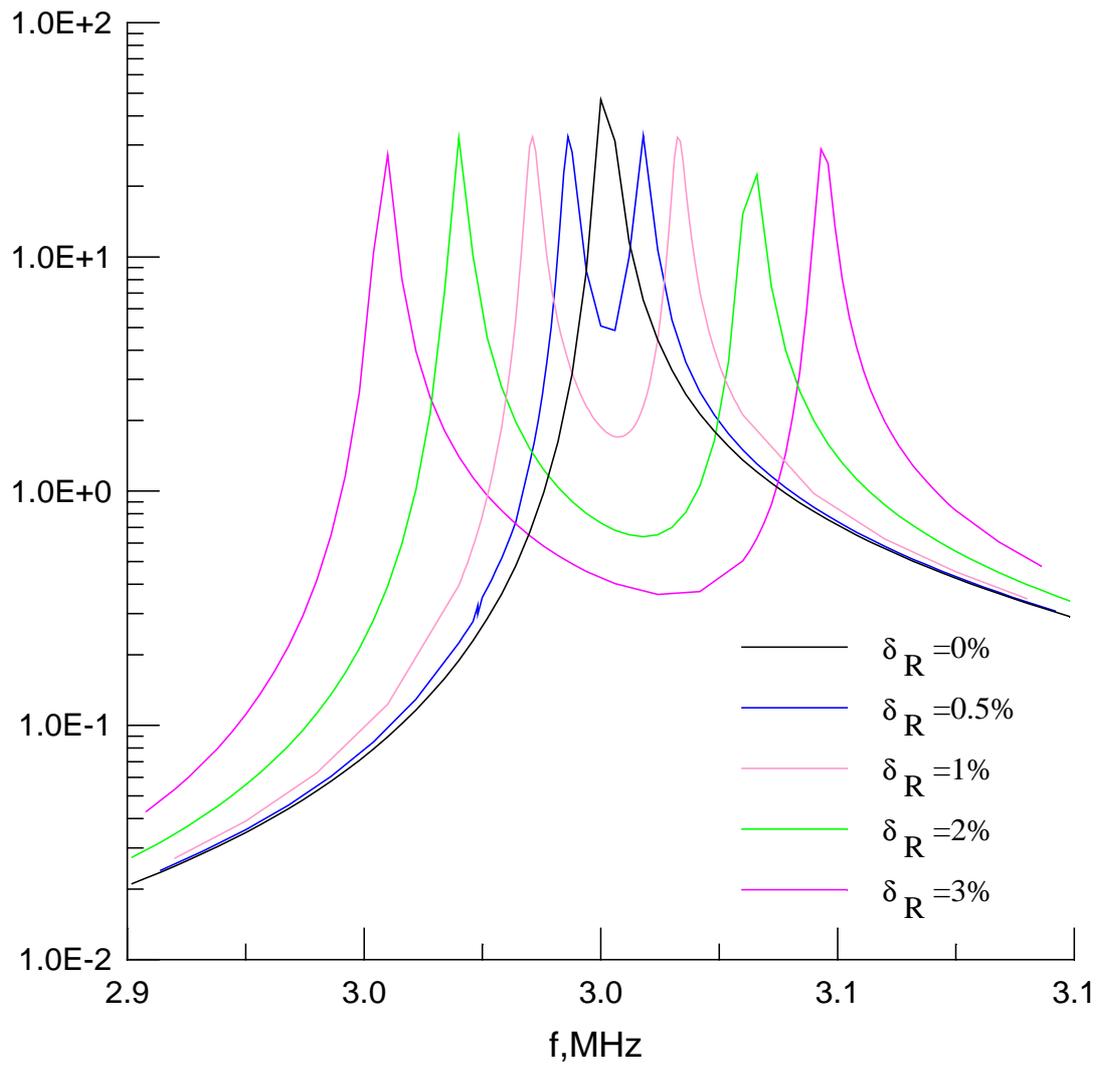

Fig.6. Dimensionless quasi-thermal noise spectrum in the ionosphere.

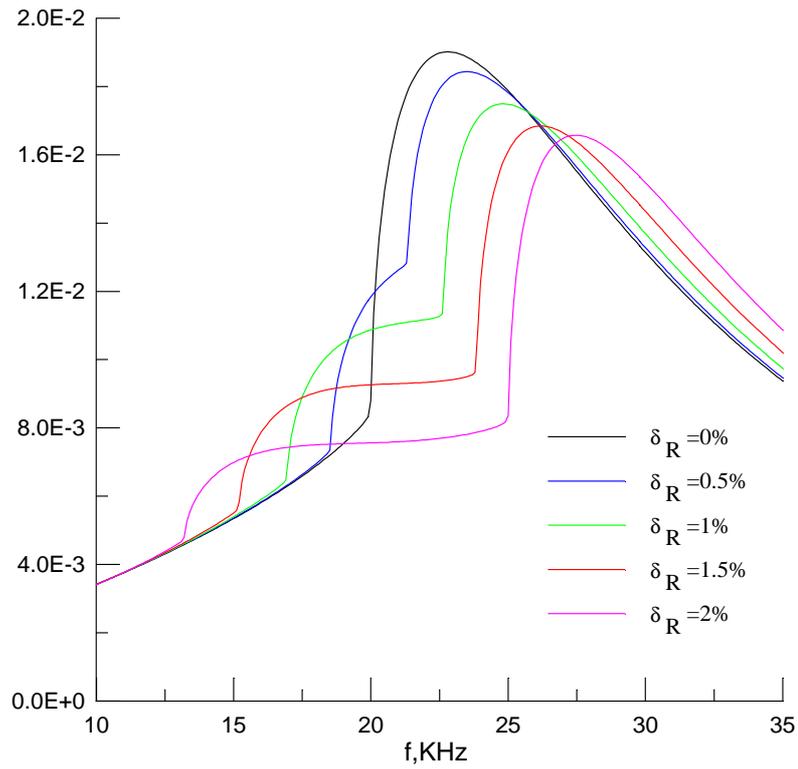

Fig.7. Dimensionless quasi-thermal noise spectrum in the solar wind; normalization scale $R = 10^5$ m.

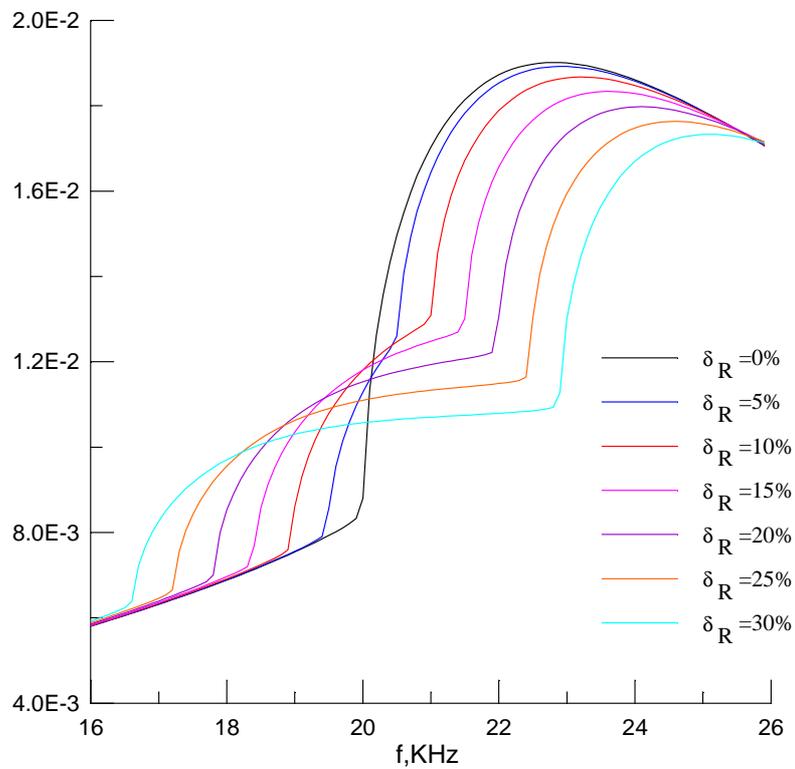

Fig.8. Dimensionless quasi-thermal noise spectrum in the solar wind; normalization scale $R = 10^8$ m.